\documentclass[journal]{IEEEtran}
\pdfoutput=1
\usepackage{url}
\usepackage{graphicx}
\usepackage{cite}
\usepackage{bm}
\usepackage{subfigure}
\usepackage{amsmath}
\usepackage{array}
\usepackage[colorlinks,linkcolor=red]{hyperref}
\usepackage{longtable}
\usepackage{amsfonts}
\usepackage{amssymb}
\usepackage{epstopdf}
\usepackage{amsthm}
\usepackage{booktabs}
\usepackage{multirow}
\usepackage{color}% marked in red
\usepackage[margin=2cm]{geometry}
\usepackage{lineno} % line number
\usepackage{dblfloatfix}
\usepackage{float}
\usepackage{booktabs}
\usepackage{tabu}
\usepackage{setspace}
\usepackage{pifont}
\usepackage{cite}
\usepackage{multicol}
\begin{document}
\title{TriDoNet: A Triple Domain Model-driven Network for CT Metal Artifact Reduction}

\author{Baoshun Shi, Ke Jiang, Shaolei Zhang, Qiusheng Lian, and Yanwei Qin
\thanks{This work was supported by the National Natural Science Foundation of China under Grants 61901406, 61871275, 61827809, 62271330, and 61971292, by the Natural Science Foundation of Hebei Province under Grant F2020203025 and F2022203030, by the Young Talent Program of Universities and Colleges in Hebei Province under Grant BJ2021044, and by the Hebei Key Laboratory Project under Grant 202250701010046.}
\thanks{B. S. Shi (Corresponding author), K. Jiang, S. L. Zhang, and Q. S. Lian are with the School of Information Science and Engineering, Yanshan University, Qinhuang Dao 066004, China. (E-mail: shibaoshun@ysu.edu.cn).\par Y. W. Qin is with the School of Mathematical Sciences, Capital Normal University, Beijing 100048, China.}}

\maketitle

\begin{abstract}
Recent deep learning-based methods have achieved promising performance for computed tomography metal artifact reduction (CTMAR). However, most of them suffer from two limitations: ($\bm i$) the domain knowledge is not fully embedded into the network training; ($\bm {ii}$) metal artifacts lack effective representation models. The aforementioned limitations leave room for further performance improvement. Against these issues, we propose a novel triple domain model-driven CTMAR network, termed as TriDoNet, whose network training exploits triple domain knowledge, i.e., the knowledge of the sinogram, CT image, and metal artifact domains. Specifically, to explore the non-local repetitive streaking patterns of metal artifacts, we encode them as an explicit tight frame sparse representation model with adaptive thresholds. Furthermore, we design a contrastive regularization (CR) built upon contrastive learning to exploit clean CT images and metal-affected images as positive and negative samples, respectively. Experimental results show that our TriDoNet can generate superior artifact-reduced CT images.
\end{abstract}

\begin{IEEEkeywords}
Metal artifact reduction, model-driven network, tight frame, triple domain network.
\end{IEEEkeywords}

\IEEEpeerreviewmaketitle
\section{Introduction}
\label{sec1}
\IEEEPARstart
{C}{omputed} tomography (CT) has become essential for clinical diagnosis and treatment plans. However, due to the metallic implants within patients, X-ray projections are incomplete, causing severe non-local streaking and star-shape artifacts in CT images \cite{9609987,8953298,10.1007/978-3-030-87231-1_11}. Reducing these structured and non-local metal artifacts is a critical problem in the field of CT \cite{9201079}. The early attempts to address this problem were the sinogram completion and iterative metal artifact reduction (MAR) algorithms. However, sinogram completion methods, such as linear interpolation (LI) \cite{Kalender1987ReductionOC} and normalized MAR (NMAR) \cite{5401721}, usually lead to secondary artifacts in restored CT images. Moreover, iterative MAR algorithms require hand-crafted regularizers, and their iteration processes are time-consuming \cite{Zhang2018ARJ}.\par
Recently, MAR methods based on deep learning have achieved promising performance. According to different domain knowledge, the existing deep learning-based MAR methods can be categorized into three classes: sinogram domain networks, image domain networks, and dual domain networks. Among them, sinogram domain networks correct metal-affected sinograms and then restore CT images by filtered back projection (FBP), which inevitably produces new artifacts \cite{8331163,8815915}. Image domain methods regard the MAR task as an image restoration task and design the MAR networks based on image domain knowledge but ignore sinogram domain knowledge \cite{2018Metal,2018Conditional}. More recent methods address this issue by embedding dual domain knowledge, such as \cite{9765584,8576532,10.1007/978-3-030-59713-9_15,10.1007/978-3-030-87231-1_24,2022arXiv220507471W}. Although current state-of-the-art (SOTA) dual domain networks have achieved high-quality CT images, less domain knowledge utilized for training and lacking effective representation models of metal artifacts are two main limitations, leaving room for further performance improvement. To mitigate these gaps, we propose a novel triple domain model-driven network. Our main contributions and innovations are summarized as follows.\par
$\bullet$ We propose the so-called TriDoNet, a triple domain unfolding network for MAR. TriDoNet ensures information interaction between the sinogram, CT image, and metal artifact domains in order to help them promote and complement each other. More precisely, we first formulate a triple domain optimization model and then solve it with an alternating iterative algorithm. The iterative steps of the algorithm are unfolded into corresponding network modules, thus the network architecture inherits the interpretability of iterative algorithms.\par
$\bullet$ We propose a supervised tight frame learning method in the metal artifact domain to represent the delicate structures of metal artifacts. A deep threshold network (DTN) is elaborated to determine spatial-varied thresholds. The learned tight frame with adaptive thresholds can capture the structure information of metal artifacts. Ablation experiments have demonstrated that the metal artifact domain sub-network using tight frame and DTN could achieve better results.\par
$\bullet$ We propose a contrastive regularization (CR) to further improve the performance of MAR networks. CR can ensure that restored CT images are pulled closer to clean CT images and pushed far away from metal-affected CT images in the representation space. Experimental results demonstrate the superior performance of TirDoNet compared with SOTA MAR methods, such as ACDNet \cite{2022arXiv220507471W} and InDuDoNet \cite{10.1007/978-3-030-87231-1_11}.
\section{The proposed methods}
\subsection{Triple domain model-driven network}
\label{sec2.1}
For an observed metal-affected CT image ${\bm y}$, it consists of two regions, namely metal region and non-metal region \cite{ Lyu2021UDuDoNetUD}. In this paper, we focus on the non-metal region. Mathematically, the decomposition model can be expressed as
\begin{equation}
	{\bm m}\odot{\bm y}={\bm m}\odot{\bm x}+{\bm m}\odot{\bm e}
	\label{eq:1}
\end{equation}
where ${\bm m}$ is the binary non-metal mask, ${\bm x}$ is the artifact-free CT image, ${\bm e}$ is the metal artifact, and $\odot$ is an element-wise product.\par
Generally, the normalized metal-affected sinogram is homogeneous \cite{10.1007/978-3-030-87231-1_11}. Therefore, we correct the normalized one. Formally, the sinogram ${\bm s}$ can be written as
\begin{equation}
	{\bm s}=\widetilde{\bm z}\odot\widetilde{\bm s}
	\label{eq:2}
\end{equation}
where $\widetilde{\bm z}$ is the normalization coefficient obtained from the prior sinogram, and $\widetilde{\bm s}$ is the normalized sinogram.\par
To explore triple domain knowledge, i.e., the knowledge of the sinogram, CT image, and metal artifact domains, a possible triple domain MAR optimization problem is formulated as
\begin{equation}
	\begin{split}
		\mathop{\min}\limits_{\widetilde{\bm s},{\bm e},{\bm x}}&||{\bm m}\odot({\bm y}-{\bm x}-{\bm e})||_2^2+\alpha||{\bm P}{\bm x}-\widetilde{\bm z}\odot\widetilde{\bm s}||_2^2\\&+\beta||(1-{\bm t}{\bm r})\odot(\widetilde{\bm z}\odot\widetilde{\bm s}-{\bm s}_{ma})||_2^2\\&+\gamma_1{R}_1(\widetilde{\bm s})+\gamma_2{R}_2({\bm e})+\gamma_3{R}_3({\bm x})
		\label{eq:3}
	\end{split}
\end{equation}
where ${\bm t}{\bm r}$ is the binary metal trace, ${\bm P}$ is the Radon transform, i.e., forward projection (FP), and ${\bm s}_{ma}$ is the metal-affected sinogram. In Eqn. (\ref{eq:3}), ${R}_1(\bullet)$, ${R}_2(\bullet)$, and ${R}_3(\bullet)$ are regularization functions that impose some desirable properties onto $\widetilde{\bm s}$, ${\bm e}$, and ${\bm x}$, respectively. Moreover, $\alpha$, $\beta$, $\gamma_1$, $\gamma_2$, and $\gamma_3$ are weight factors.\par
We explore the sparsity of metal artifacts over the learned tight frame ${\bm W}\in\mathbb{R}^{M \times N}(M \geq N)$ to formulate the regularizer of metal artifacts, i.e., ${R}_2({\bm e})=||{\bm W}{\bm e}||_1$. Here, $||\bullet||_1$ represents the ${l}_1$ norm. The triple domain reconstruction problem defined in Eqn. (\ref{eq:3}) can be recast as
\begin{equation}
	\begin{split}
		\mathop{\min}\limits_{\widetilde{\bm s},{\bm e},{\bm x}}&||{\bm m}\odot({\bm y}-{\bm x}-{\bm e})||_2^2+\alpha||{\bm P}{\bm x}-\widetilde{\bm z}\odot\widetilde{\bm s}||_2^2\\&+\beta||(1-{\bm t}{\bm r})\odot(\widetilde{\bm z}\odot\widetilde{\bm s}-{\bm s}_{ma})||_2^2\\&+\gamma_1{R}_1(\widetilde{\bm s})+\gamma_2||{\bm W}{\bm e}||_1+\gamma_3{R}_3({\bm x}).
		\label{eq:4}
	\end{split}
\end{equation}\par
We exploit an alternating iteration method to solve the triple domain MAR optimization problem defined in Eqn. (\ref{eq:4}), and proximal operators are utilized to solve each sub-problem. At the $({t}+1)$-th iteration, ${\widetilde{\bm s}}$, ${\bm e}$, and ${\bm x}$ are alternately updated as follows.\par
{\textbf{Updating}} $\widetilde{\bm s}$: Given ${\bm e}^{(t)}$ and ${\bm x}^{(t)}$, the sub-problem of updating $\widetilde{\bm s}$ is derived as
\begin{equation}
	\begin{split}
		&\mathop{\min}\limits_{\widetilde{\bm s}}\alpha||{\bm P}{\bm x}^{(t)}-\widetilde{\bm z}\odot\widetilde{\bm s}||_2^2\\&+\beta||(1-{\bm t}{\bm r})\odot(\widetilde{\bm z}\odot\widetilde{\bm s}-{\bm s}_{ma})||_2^2+\gamma_1{R}_1(\widetilde{\bm s}).
		\label{eq:5}
	\end{split}
\end{equation}
The quadratic approximation of the sub-problem defined in Eqn. (\ref{eq:5}) is
\begin{equation}
	\begin{split}
		\mathop{\min}\limits_{\widetilde{\bm s}}&\,{g}_1({\widetilde{\bm s}}^{(t)})+\langle{\widetilde{\bm s}}-{\widetilde{\bm s}^{(t)}},\bigtriangledown{g}_1({\widetilde{\bm s}})\rangle\\&+\frac{1}{2\eta_1}||{\widetilde{\bm s}}-{\widetilde{\bm s}}^{(t)}||_2^2+\gamma_1{R}_1(\widetilde{\bm s})
		\label{eq:6}
	\end{split}
\end{equation}
where ${g}_1(\widetilde{\bm s}^{(t)})=\alpha||{\bm P}{\bm x}^{(t)}-\widetilde{\bm z}\odot\widetilde{\bm s}^{(t)}||_2^2+\beta||(1-{\bm t}{\bm r})\odot(\widetilde{\bm z}\odot\widetilde{\bm s}^{(t)}-{\bm s}_{ma})||_2^2$, and $\eta_1$ is the stepsize. Equation (\ref{eq:6}) can be equivalently written as
\begin{equation}
	\begin{split}
		\mathop{\min}\limits_{\widetilde{\bm s}}\frac{1}{2}||\widetilde{\bm s}-(\widetilde{\bm s}^{(t)}-\eta_1\bigtriangledown{g}_1({\widetilde{\bm s}}^{(t)}))||_2^2+\gamma_1\eta_1{R}_1(\widetilde{\bm s})
		\label{eq:7}
	\end{split}
\end{equation}
where $\bigtriangledown{g}_1({\widetilde{\bm s}^{(t)}})=\alpha\widetilde{\bm z}\odot(\widetilde{\bm z}\odot\widetilde{\bm s}^{(t)}-{\bm P}{\bm x}^{(t)})+\beta(1-{\bm t}{\bm r})\odot\widetilde{\bm z}\odot(\widetilde{\bm z}\odot\widetilde{\bm s}^{(t)}-{\bm s}_{ma})$.
Using the proximal gradient algorithm, the update rule of ${\widetilde{\bm s}}$ can be described as
\begin{equation}
	\begin{split}
		{\widetilde{\bm s}}^{(t+1)}&\triangleq{prox}_{\gamma_1\eta_1{R}_1}({\widetilde{\bm s}}^{(t+0.5)}).
		\label{eq:8}
	\end{split}
\end{equation}
where ${prox}_{\gamma_1\eta_1{R}_1}(\bullet)$ is a proximal operator related to ${R}_1(\bullet)$. Given ${\widetilde{\bm s}}^{(t+0.5)}=\widetilde{\bm s}^{(t)}-\eta_1\bigtriangledown{g}_1({\widetilde{\bm s}}^{(t)})$, we have
\begin{equation}
	\begin{split}
		{\widetilde{\bm s}}^{(t+0.5)}&={\widetilde{\bm s}}^{(t)}-\eta_1\alpha\widetilde{\bm z}\odot(\widetilde{\bm z}\odot\widetilde{\bm s}^{(t)}-{\bm P}{\bm x}^{(t)})\\&-\eta_1\beta(1-{\bm t}{\bm r})\odot\widetilde{\bm z}\odot(\widetilde{\bm z}\odot\widetilde{\bm s}^{(t)}-{\bm s}_{ma}).
		\label{eq:9}
	\end{split}
\end{equation}\par
{\textbf{Updating}} ${\bm e}$: Fix $\widetilde{\bm s}^{(t+1)}$ and ${\bm x}^{(t)}$, we can update ${\bm e}$ by solving the following sub-problem
\begin{equation}
	\begin{split}
		\mathop{\min}\limits_{{\bm e}}||{\bm m}\odot({\bm y}-{\bm x}^{(t)}-{\bm e})||_2^2+\gamma_2||{\bm W}{\bm e}||_1.
		\label{eq:10}
	\end{split}
\end{equation}
Similarly, the updating rule of ${\bm e}$ is written as
\begin{equation}
	\begin{split}
		{{\bm e}}^{(t+1)}&\triangleq{prox}_{\gamma_2\eta_2{R}_2}({\bm e}^{(t+0.5)})
		\label{eq:11}
	\end{split}
\end{equation}
where
\begin{equation}
	\begin{split}
		{{\bm e}}^{(t+0.5)}=(1-\eta_2{\bm m})\odot{\bm e}^{(t)}+\eta_2{\bm m}\odot({\bm y}-{\bm x}^{(t)}).
		\label{eq:12}
	\end{split}
\end{equation} 
The tight frame ${\bm W}$ satisfies tight property $||{\bm W}{\bm e}||_2^2=||{\bm e}||_2^2$, thus Eqn. (\ref{eq:11}) can be expressed as
\begin{equation}
	\begin{split}
		{\bm e}^{(t+1)}={\bm W}^{\rm T}{soft}({\bm W}{\bm e}^{(t+0.5)},\varepsilon)
		\label{eq:13}
	\end{split}
\end{equation}
where $soft(u,\varepsilon)=sign(u)max(|u|-\varepsilon,0)$ is the soft threshold function, and $\varepsilon=\gamma_2\eta_2$ is the threshold. In this paper, we utilize an elaborated DTN to adaptively determine the thresholds from the coefficients of metal artifacts, i.e., $\varepsilon=f({\bm W}{\bm e}^{(t+0.5)})$. Here, ${f}(\bullet)$ is the so-called DTN, whose network architecture is described in Sec. \ref{sec2.2}.\par
\begin{figure*}[htbp]
	%	\vspace{-0.7cm}  %调整图片与上文的垂直距离	
	%	\setlength{\abovecaptionskip}{-0.0cm}   %调整图片标题与图距离	
	%	\setlength{\belowcaptionskip}{-0.8cm}   %调整图片标题与下文距离
	\centering
	{\includegraphics[width=17.3cm]{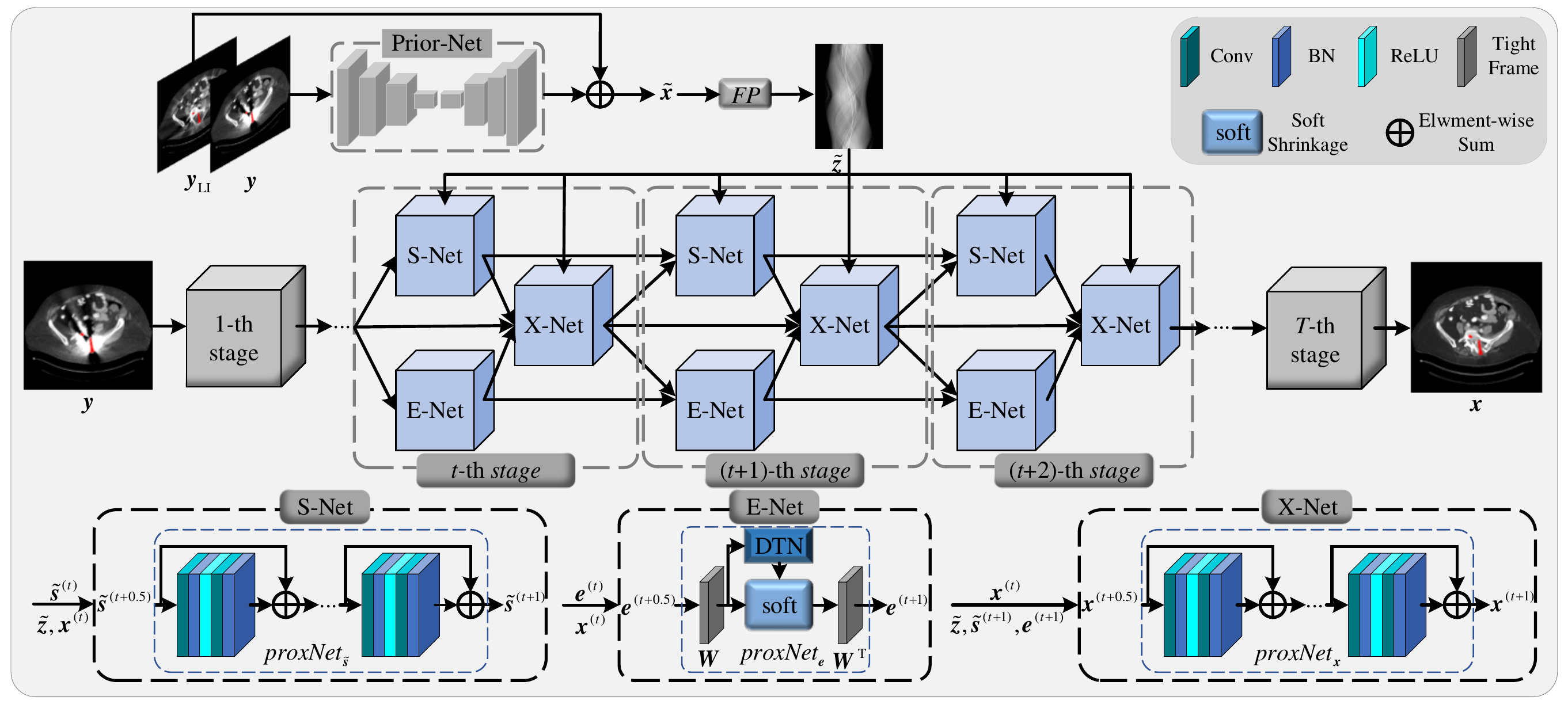}}
	\caption{The proposed TriDoNet architecture. The detailed structure at the $(t+1)$-th stage where $\widetilde{\bm s}^{(t+1)}$, ${\bm e}^{(t+1)}$, and ${\bm x}^{(t+1)}$ are updated by S-Net, E-Net, and X-Net, respectively, based on Eqn. (\ref{eq:8}), Eqn. (\ref{eq:11}), and Eqn. (\ref{eq:15}). Specifically, S-Net and X-Net contain several residual blocks, and E-Net contains DTN to determine the thresholds. ${proxNet}_{\widetilde{\bm s}}$, ${proxNet}_{\bm e}$, and ${proxNet}_{\bm x}$ denote the proximal networks of the variables ${\widetilde{\bm s}}$, ${\bm e}$, and ${\bm x}$, respectively. The triple domain information is passed between and within each stage, enabling the features of triple domains to interact and promote each other.}
	\label{Fig1}
\end{figure*}
{\textbf{Updating}} ${\bm x}$: Given $\widetilde{\bm s}^{(t+1)}$ and ${\bm e}^{(t+1)}$, ${\bm x}$ is updated by solving the following sub-problem
\begin{equation}
	\begin{split}
		\mathop{\min}\limits_{\bm x}&||{\bm m}\odot({\bm y}-{\bm x}-{\bm e}^{(t+1)})||_2^2\\&+\alpha||{\bm P}{\bm x}-\widetilde{\bm z}\odot\widetilde{\bm s}^{(t+1)}||_2^2+\gamma_3{R}_3({\bm x}).
		\label{eq:14}
	\end{split}
\end{equation}
Similarly, the updating rule of ${\bm x}$ is written as
\begin{equation}
	\begin{split}
		{{\bm x}}^{(t+1)}&\triangleq{prox}_{\gamma_3\eta_3{R}_3}({\bm x}^{(t+0.5)})
		\label{eq:15}
	\end{split}
\end{equation}
where
\begin{align}
	&{\bm x}^{(t+0.5)}=(1-\eta_3){\bm m}\odot{\bm x}^{(t)}\\&+\eta_3{\bm m}\odot({\bm y}-{\bm e}^{(t+1)})-\alpha\eta_3{\bm P}^{\rm T}({\bm P}{\bm x}^{(t)}-\widetilde{\bm z}\odot\widetilde{\bm s}^{(t+1)}).\nonumber		
	\label{eq:16}
\end{align}
\par
\subsection{Network architecture and loss function}
\label{sec2.2}
In this section, we unroll the proposed iterative algorithm in Sec. \ref{sec2.1} into a MAR network termed TriDoNet. As illustrated in Fig. \ref{Fig1}, the proposed TriDoNet consists of $T$ stages corresponding to $T$ iterations of the iterative algorithm, and each stage contains S-Net, E-Net, and X-Net. The figure shows how the data flow within and between each stage, and this information interaction facilitates the reconstruction of favorable details in triple domains. Moreover, Prior-Net is utilized to produce the normalization coefficient $\widetilde{\bm z}$.\par
{\textbf {Prior-Net}}: The prior network in Fig. \ref{Fig1} is utilized to recover the preliminary corrected CT image $\widetilde{\bm x}$ from the linear interpolation corrected CT image ${\bm y}_{\rm LI}$ and the metal-affected CT image ${\bm y}$. Then, the prior image $\widetilde{\bm x}$ passes through the FP to obtain the prior sinogram $\widetilde{\bm z}$.\par
{\textbf {S-Net}}: At the $(t+1)$-th stage, with $\widetilde{\bm z}$ generated by Prior-net, $\widetilde{\bm s}^{(t+0.5)}$ is obtained by Eqn. (\ref{eq:9}). We can get $\widetilde{\bm s}^{(t+1)}={proxNet}_{\widetilde{\bm s}}(\widetilde{\bm s}^{(t+0.5)})$, where ${proxNet}_{\widetilde{\bm s}}(\bullet)$ consists of several [Conv$+$BN$+$ReLU$+$Conv$+$BN$+$Skip Connection] residual blocks.\par
{\textbf {E-Net}}: Given $\widetilde{\bm s}^{(t+1)}$ and ${\bm x}^{(t+1)}$, ${\bm e}^{(t+0.5)}$ is obtained and fed to ${proxNet}_{\bm e}(\bullet)$ at the $(t+1)$-th stage. The update rule of ${\bm e}$ is ${\bm e}^{(t+1)}={proxNet}_{\bm e}({\bm e}^{(t+0.5)})$, where ${proxNet}_{\bm e}(\bullet)$ is a tight frame network. To adaptively extract the thresholds from the inputs, we explore the DTN built by six convolution layers. Additionally, ${\bm W}$ is learned by back-propagation in the training phase.\par
{\textbf {X-Net}}: Similarly, at the $(t+1)$-th stage, given $\widetilde{\bm s}^{(t+1)}$, ${\bm e}^{(t+1)}$, and $\widetilde{\bm z}$, the CT images can be updated by ${\bm x}^{(t+1)}={proxNet}_{\bm x}({\bm x}^{(t+0.5)})$, where ${proxNet}_{\bm x}(\bullet)$ consists of several [Conv$+$BN$+$ReLU$+$Conv$+$BN$+$Skip Connection] residual blocks.\par
{\textbf {Loss Function}}: To utilize triple domain knowledge, the triple domain loss function is designed to embed triple domain knowledge into the network training. With supervision on the sinogram $\widetilde{\bm s}^{(t)}$, the metal artifact ${\bm e}^{(t)}$, and CT image ${\bm x}^{(t)}$ at every stage, the triple domain loss function is defined as
\begin{align}
	&L=\omega_1||{\bm W}^{{\rm T}}{\bm W}-{\bm I}||_2^2\\
	&+\sum_{t=0}^{T}\{\omega_2||{\bm m}\odot({\bm x}-{\bm x}^{(t)})||_2^2+\omega_3{||{\bm m}\odot({\bm x}-{\bm x}^{(t)})||_1}\nonumber\\
	&+\omega_4||{\bm m}\odot({\bm y}-{\bm x}-{\bm e}^{(t)})||_1+\omega_5||\widetilde{\bm z}\odot\widetilde{\bm s}^{(t)}-{\bm s}_{gt}||_2^2\nonumber
	\label{eq:17}\}
\end{align}
\begin{table*}[h]
	%	\vspace{-0.3cm}
	\centering
	\caption{Quantitative evaluations for different MAR methods on DeepLesion data. We report the average PSNR (dB)/SSIM values of the testing dataset for each case. The best results are highlighted in bold.}
	\footnotesize{\tabcolsep=10pt}
	%	\vspace{-0.1cm}  %调整图片与上文的垂直距
	\begin{tabular}{l|ccccc|c}
		\toprule[1.3pt]
		{\bf Method} & \multicolumn{2}{c}{ \bf Large Metal} & $\longrightarrow$ & \multicolumn{2}{c}{\bf Small Metal}  \vline& {\bf Average} \\
		\midrule
		{Input}  &21.60 / 0.5442 &24.72 / 0.6723 &29.34 / 0.7185&30.02 / 0.7541 & 31.60 / 0.7571   &27.46 / 0.6892\\
		{BHC} &24.57 / 0.6629    &26.70 / 0.7327 &29.80 / 0.7917&30.26 / 0.8152 &30.72 / 0.8183    &28.41 / 0.7642\\
		{LI} &31.30 / 0.9031     &32.95 / 0.9295 &36.41 / 0.9600  &37.46 / 0.9649 & 38.14 / 0.9723   &35.20 / 0.9460\\
		{NMAR} &31.75 / 0.9529   &33.83 / 0.9484 &36.91 / 0.9641  &37.65 / 0.9635 & 38.70 / 0.9782   &35.77 / 0.9553\\
		{CNNMAR} &33.19 / 0.9529 &35.76 / 0.9700 &38.69 / 0.9817  &39.13 / 0.9831 & 39.48 / 0.9855   &37.25 / 0.9746\\
		{DuDoNet} &33.52 / 0.9603&35.36 / 0.9708 &40.49 / 0.9812  &42.04 / 0.9830 &42.60 / 0.9841    &38.80 / 0.9759\\
		{InDuDoNet} &37.70 / 0.9708  &38.83 / 0.9793 &42.95 / 0.9858&45.28 / 0.9877&45.50 / 0.9881 &42.05 / 0.9824\\
		{ACDNet} &37.82 / 0.9766     &39.60 / 0.9828 &43.85 / 0.9879&45.46 / 0.9892&45.81 / 0.9896 &42.51 / 0.9852\\
		{TriDoNet} &{40.71 / 0.9835}&{37.70 / 0.9837}&{44.09 / 0.9892}&{46.55 / 0.9906}&{47.03 / 0.9911}&\textbf{43.22 / 0.9876}\\
		\bottomrule[1.3pt]
		%\toprule   
	\end{tabular}
	\label{tab1}
	%	\vspace{-0.8cm}
\end{table*}
where ${\bm s}_{gt}$ is the metal-free sinogram. The first term is the tight frame loss, the second and third terms are utilized for the image domain loss, the fourth term is the metal artifact domain loss, and the fifth term is the sinogram domain loss.\par
Inspired by contrastive learning \cite{9578448}, we fuse contrastive regularization to the triple domain loss to improve the quality of restored CT images. We define clean images ${\bm x}$ as ``positive'' samples and metal-affected images ${\bm y}$ as ``negative'' samples. We select common intermediate features from the same fixed pre-training model $G(\bullet)$ for the latent feature, e.g., VGG-19. Precisely, the formulated contrastive loss is defined as
\begin{equation}
	\begin{split}
		&CR({\bm x},{\bm x}^{(T)},{\bm y})={\rho}\sum_{i=1}^{n}\frac{||G_i(\bm x^{(T)})-G_i(\bm x)||_1}{||G_i(\bm x^{(T)})-G_i({\bm y})||_1}
		\label{eq:18}
	\end{split}
\end{equation}
where $G_i$, $i=1,2,...n$ extracts the $i$-th hidden features by using the fixed pre-trained VGG model, and $\rho$ is a hyperparameter.\par
\section{Experiments and results}
\label{sec3}
\textbf{Dataset}: Following the simulation procedure in \cite{9201079,8788607}, we select 1200 clean CT images from DeepLesion dataset \cite{8579063} and 100 metal masks from \cite{8331163} to synthesize metal artifacts at random. All CT images are resized to $416\times416$. The size of the generated sinogram is $641\times640$. We choose 1000 images and 90 metal masks for training and 10 metal masks together with 200 images for testing. The details of the experiment are provided in the supplementary material.\par
\begin{figure}[h]
	%	\vspace{-0.5cm}  %调整图片与上文的垂直距离	
	%	\setlength{\abovecaptionskip}{0.1cm}   %调整图片标题与图距离	
	%	\setlength{\belowcaptionskip}{-0.5cm}   %调整图片标题与下文距离
	\centering
	{\includegraphics[width=8.3cm]{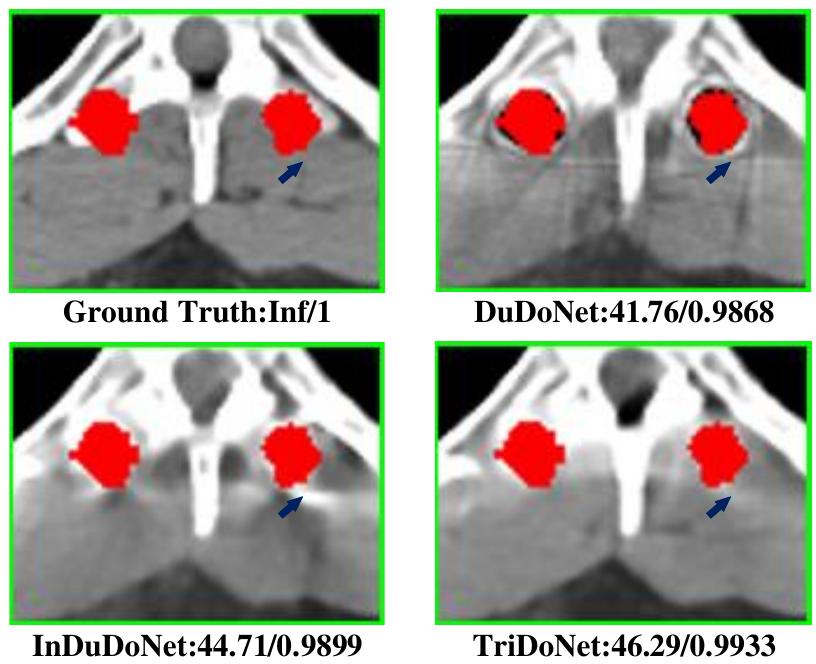}}
	\caption{Comparison for medium metallic implants from DeepLesion data. PSNR (dB)/SSIM is listed for reference. The red pixels stand for metallic implants. The window width is 450HU, and the window level is 50 HU.}
	\label{Fig2}
\end{figure}

\textbf{Performance Evaluation}: To demonstrate the superiority of our TriDoNet, we compare the MAR performance of BHC \cite{Verburg_2012}, LI \cite{Kalender1987ReductionOC}, NMAR \cite{5401721}, CNNMAR \cite{8331163}, DuDoNet \cite{8953298}, InDuDoNet \cite{10.1007/978-3-030-87231-1_11}, ACDNet \cite{2022arXiv220507471W}, and the proposed TriDoNet. We employ peak signal-to-noise ratio (PSNR) and structural similarity index measure (SSIM) for quantitative evaluation. Table \ref{tab1} summarizes the average PSNR and SSIM values on varying metal sizes. It can be observed that the proposed TriDoNet can achieve the highest average PSNR and SSIM values. Moreover, the average PSNR value of TriDoNet is nearly 0.71 dB and 1.17 dB higher than those of ACDNet and InDuDoNet, respectively. High-quality CT images restored by our TriDoNet are mainly attributed to the utilization of triple domain knowledge and the tight frame network. Figure \ref{Fig2} presents the MAR visual comparison. From the visual perspective, the proposed TriDoNet recovers evident tissue details finely and removes most metal artifacts.\par
\textbf{Ablation Study}: To demonstrate the effectiveness of TriDoNet, we compare the performance of different components. Under the same conditions, we compare single domain networks and dual domain networks for the MAR task. We remove CR to verify the effectiveness of contrastive learning and replace the threshold generated by DTN with a learnable parameter to demonstrate the effectiveness of the DTN. Table \ref{tab2} shows that X-Net, S-Net, and E-net play their respective roles in processing each domain for the MAR task. Meanwhile, experiments demonstrate the effectiveness of CR and DTN in our model.\par
\begin{table}[h]
	%	\vspace{-0.2cm}
	\centering
	\caption{Average PSNR (dB)/SSIM values achieved by TriDoNet with different components on synthesized data. The variation represents the gap with our algorithm, and the blue arrow indicates the decreased value.}
	\footnotesize{\tabcolsep=10pt}
	%	\vspace{-0.2cm}  %调整图片与上文的垂直距
	\begin{tabular}{ccccc|c|c}
		\toprule[1.3pt]
		\makebox[0.02\textwidth][c]{$\text{X-Net}$} & \makebox[0.02\textwidth][c]{$\text{S-Net}$} & \makebox[0.02\textwidth][c]{$\text{E-Net}$}& \makebox[0.02\textwidth][c]{$\text{CR}$} & \makebox[0.02\textwidth][c]{$\text{DTN}$} & \makebox[0.1\textwidth][c]{$\text{PSNR/SSIM}$} & \makebox[0.03\textwidth][c]{$\text{variation}$} 	\\
		%		$\text{X-Net}$ & $\text{S-Net}$ & $\text{E-Net}$ & $\text{CR}$ & $\text{DTN}$ & $\text{PSNR/SSIM} $ & $\text{variation} $ \\
		\midrule
		\ding{55}&\ding{51}&\ding{55}&\ding{51}&\ding{51}  &37.73/0.8899 &\textcolor{blue}{$\downarrow$}5.49/0.0977\\
		\ding{51}&\ding{55}&\ding{55}&\ding{51}&\ding{51}  &40.16/0.9797 &\textcolor{blue}{$\downarrow$}3.06/0.0079\\
		\ding{51}&\ding{51}&\ding{55}&\ding{51}&\ding{51}  &42.29/0.9847 &\textcolor{blue}{$\downarrow$}0.93/0.0029\\
		\ding{51}&\ding{55}&\ding{51}&\ding{51}&\ding{51}  &42.83/0.9868 &\textcolor{blue}{$\downarrow$}0.39/0.0008\\
		\ding{51}&\ding{51}&\ding{51}&\ding{55}&\ding{51}  &42.69/0.9872 &\textcolor{blue}{$\downarrow$}0.53/0.0004\\
		\ding{51}&\ding{51}&\ding{51}&\ding{51}&\ding{55}  &43.11/0.9858 &\textcolor{blue}{$\downarrow$}0.11/0.0018\\ 
		\ding{51}&\ding{51}&\ding{51}&\ding{51}&\ding{51}  &43.22/0.9876 &0/0\\
		\bottomrule[1.3pt]
	\end{tabular}
	\label{tab2}
	%	\vspace{-0.3cm}
\end{table}
\section{Conclusion}
In this paper, we embedded triple domain knowledge into a deep unfolding MAR framework and elaborated a tight frame network with adaptive thresholds to restore the metal artifacts. We also explored a contrastive loss to improve performance. Experiments show the effectiveness of the proposed triple domain model-driven MAR network.\par

\vfill\pagebreak
\bibliographystyle{IEEEbib}
\bibliography{ref}
\end{document}

%\markboth{Journal of \LaTeX\ Class Files, Vol. 14, No. 8, August 2022}

\section{Training Details}
\label{sec:Training Details}
TriDoNet is implemented using PyTorch. We use the ADAM algorithm as the optimizer with $(\beta_1,\beta_2)=(0.5,0.999)$ whose initial learning rate is $2\times10^{-4}$ and decreased by a factor of 0.5 every 20 epochs. The model is trained on NVIDIA 3090Ti for 70 epochs. The parameters $\beta$, $\eta_1$, $\eta_2$, and $\eta_3$ are automatically learned from training samples in an end-to-end manner. Heuristically, the parameters are set to $\alpha$ = 1, $\rho$ = 0.1, $\mu_{t}$ $(t=0,1,...,T-1)$ = 0.1, and $\mu_{T}$ = 1. Moreover, $\omega_1$, $\omega_2$, $\omega_3$, $\omega_4$, and $\omega_5$ are hyperparameters to balance the weight of different loss items and we empirically set them to $0.001$, $\mu_{t}$, $5\times10^{-4}\mu_{t}$, $5\times10^{-4}\mu_{t}$, and $0.01\mu_{t}$, respectively.\par
On synthesized Deeplesion data, we select 200 CT images from 12 patients and 10 masks of different sizes to synthesize the test data. The dimensions of the metallic implants of the test data are: [2061, 890, 881, 451, 254, 124, 118, 112, 53, 35] in pixels. We compare TriDoNet with existing metal artifact reduction (MAR) methods, including BHC, LI, NMAR, CNNMAR, DuDoNet, InDuDoNet, and ACDNet. For the benchmark algorithms, we use the official codes, except for DuDoNet, we re-implement it since there is no official code.\par
\section{More Results}
\label{sec:More Experiments}
\begin{figure*}[htbp]
%	\vspace{0.3cm}  %调整图片与上文的垂直距离	
%	\setlength{\abovecaptionskip}{-0.0cm}   %调整图片标题与图距离	
%	\setlength{\belowcaptionskip}{-0.5cm}   %调整图片标题与下文距离
	\centering
	\centering
	{\includegraphics[width=17.3cm]{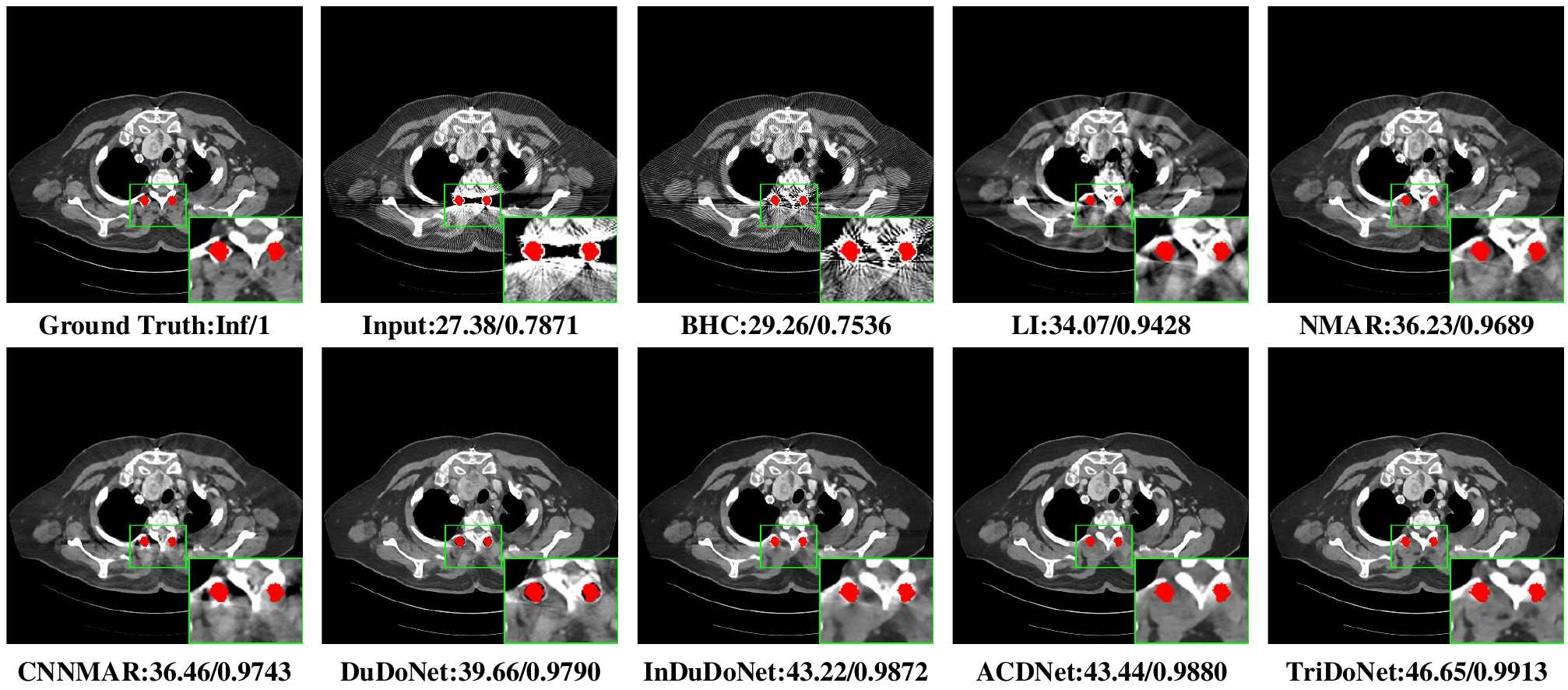}}
	\caption{Comparison for medium metallic implants from synthesized data. PSNR(dB)/SSIM is listed for reference. The magnified details are shown in the green boxes. Red pixels represent metallic implants. The window width is 450HU and the window level is 50 HU.}
	\label{fig1}
\end{figure*}

Figure \ref{fig1} shows the visual comparison between our TriDoNet and the benchmark algorithms. For the convenience of observation, we enlarge the vicinity of metallic implants to display the recovery details. At the bottom of each image, we list the PSNR and SSIM values of the corresponding methods. In terms of visual quality, the images recovered by the InDuDoNet and ACDNet algorithms are blurring, and those achieved by the BHC, LI, NMAR, CNNMAR, and DuDoNet algorithms contain many artifacts. The images recovered by our proposed MAR method are better than those achieved by the benchmark MAR algorithms in terms of preserving details and reducing metal artifacts.
\par
\vfill\pagebreak
\bibliographystyle{IEEEbib}
%\bibliography{refs}